\documentclass[runningheads]{llncs}
\usepackage[T1]{fontenc}
\usepackage{graphicx}
\usepackage{needspace}
\usepackage{amsmath}
\usepackage{amssymb}
\usepackage{subcaption}
\usepackage{float}
\usepackage[table]{xcolor}

\begin{document}
\title{Integrating Deep Learning with Fundus and Optical Coherence Tomography for Cardiovascular Disease Prediction}
\titlerunning{Integrating DL with Fundus and OCT for CVD Prediction}
\author{Cynthia Maldonado-Garcia\inst{1}\orcidID{0000-0002-3179-7070} \and
\and Arezoo Zakeri\inst{2}\orcidID{0000-0002-6011-2333}
\and Alejandro F Frangi*\inst{2,3,4,5}\orcidID{0000-0002-2675-528X}
\and Nishant Ravikumar*\inst{1}\orcidID{0000-0003-0134-107X}
}
\institute{Centre for Computational Imaging and Simulation Technologies in Biomedicine, School of Computing, University of Leeds, Leeds, UK \and
Division of Informatics, Imaging, and Data Sciences, School of Health Sciences, Faculty of Biology, Medicine, and Health, University of Manchester, Manchester, UK \and School of Computer Science, Faculty of Science and Engineering, University of Manchester, Kilburn Building, Manchester, UK \and Christabel Pankhurst Institute, University of Manchester, Manchester, UK \and NIHR Manchester Biomedical Research Centre, Manchester Academic Health Science Centre, Manchester, UK\\
\email{scclmg@leeds.ac.uk, arezoo.zakeri@manchester.ac.uk, N.Ravikumar@leeds.ac.uk, alejandro.frangi@manchester.ac.uk}\\
$*$ indicates joint last authors}

\authorrunning{{Maldonado-García} et al.}

\maketitle              
\begin{abstract}
Early identification of patients at risk of cardiovascular diseases (CVD) is crucial for effective preventive care, reducing healthcare burden, and improving patients' quality of life. This study demonstrates the potential of retinal optical coherence tomography (OCT) imaging in conjunction with fundus photographs for identifying future adverse cardiac events. We utilized data from 977 patients who experienced CVD within a 5-year interval post-image acquisition, alongside 1,877 control participants without CVD, totaling 2,854 subjects. We propose a novel binary classification network based on a Multi-channel Variational Autoencoder (MCVAE), which learns a latent embedding of patients' fundus and OCT images to classify individuals into two groups: those likely to develop CVD in the future and those who are not. Our model, trained on both imaging modalities, achieved promising results (AUROC 0.78 ± 0.02, accuracy 0.68 ± 0.002, precision 0.74 ± 0.02, sensitivity 0.73 ± 0.02, and specificity 0.68 ± 0.01), demonstrating its efficacy in identifying patients at risk of future CVD events based on their retinal images. This study highlights the potential of retinal OCT imaging and fundus photographs as cost-effective, non-invasive alternatives for predicting cardiovascular disease risk. The widespread availability of these imaging techniques in optometry practices and hospitals further enhances their potential for large-scale CVD risk screening. Our findings contribute to the development of standardized, accessible methods for early CVD risk identification, potentially improving preventive care strategies and patient outcomes.

\end{abstract}

\section{Introduction}
Cardiovascular diseases (CVD) have maintained their status as the primary global cause of mortality for several decades. Recent epidemiological data from 2021 indicates that CVD account for one-third of global deaths, with up to 80\% of premature cardiovascular conditions being potentially preventable \cite{world-heart}. A significant health disparity is evident, with approximately 80\% of CVD-related fatalities occurring in low- and middle-income countries, while advancements in cardiovascular health are predominantly observed in high-income nations. This inequity underscores an urgent need for targeted interventions and global health initiatives.

The intricate nature of biological tissues, organs, and disease processes necessitates the utilization of multi-modal imaging techniques within the medical community to accurately characterize disease phenotypes and extract clinically relevant quantitative information \cite{Rajiah2019BandsIT,Govindarajan2005TheCS}. Recent advances in artificial intelligence (AI) methodologies for computational image analysis and the extraction of disease phenotypes across multiple modalities have demonstrated promise in elucidating the underlying anatomical and physiological alterations associated with CVD \cite{Amal2022UseOM,Milosevic2024ApplicationsOA}. Consequently, the development of sophisticated statistical and machine learning techniques for effective analysis and integration of these diverse data sources is imperative for enhancing the quality of patient care.

In the domain of AI, several contemporary studies have explored multi-modality approaches for cardiovascular diseases, incorporating various types of medical data. These include imaging techniques such as magnetic resonance imaging (MRI), cardiac computed tomography (CT) scans, retinal imaging (fundus photographs and optical coherence tomography (OCT)), clinical data from electronic health records (EHR), and genetic data \cite{Milosevic2024ApplicationsOA}. These investigations have demonstrated improvements in assessing CVD risk early on and understanding the systemic changes that are indicators of increased CVD risk \cite{Mohsen2023ArtificialIM,Li2022MultiModalityCI}.

Retinal imaging studies have demonstrated significant potential in CVD prediction, a finding of particular interest due to the accessibility of such images during routine ophthalmic or optometric examinations. This accessibility suggests the possibility of cardiovascular health screening as an ancillary benefit of standard ocular assessments. While fundus imaging has been the subject of more extensive research \cite{DiazPinto2022PredictingMI,Poplin2017PredictionOC}, recent investigations have explored the predictive capabilities of OCT, further enhancing predictive accuracy \cite{Garca2024PredictingRO}. However, the literature reveals a paucity of studies employing both modalities concurrently for CVD prediction. Existing research often fails to fully exploit the potential of combined data or to conduct comprehensive comparative analyses of the efficacy of dual-modality approaches versus single-modality methods \cite{Zhou2023AFM,Huang2023AIintegratedOI}.

This study presents a novel predictive model that integrates both OCT and fundus retinal imaging data, by employing a multichannel variational autoencoder and a transformer network classifier. The proposed predictive model effectively combines multi-modal retinal imaging data to identify individuals at risk of developing cardiovascular diseases within five years of image acquisition. This approach aims to enhance our understanding of the systemic changes that contribute to an increased risk of cardiovascular diseases in patients.
\section{Data and Methodology}
\subsection{Data}
The UK Biobank is a valuable resource containing a vast array of health-related data from over 500,000 participants across the United Kingdom. This extensive dataset encompasses genetic information, demographic characteristics, clinical measurements, lifestyle factors, and medical imaging data. For the present study, we utilized retinal OCT and fundus imaging data obtained from the UK Biobank, which were acquired using the Topcon 3D OCT 1000 Mark 2 system (45° field-of-view, centered to include both the optic disc and macula). We focused on retinal imaging of the left eye, as it has been shown to have better quality compared to the right eye. During the baseline visit (2006-2010), a total of 67,656 participants underwent retinal imaging for both modalities. Quality control exclusion criteria for OCT and fundus images were applied in accordance with established protocols \cite{Zekavat2023InsightsIH,Fu2019EvaluationOR}. Following rigorous quality assessment, 43,097 participants with high-quality OCT images and 41,271 participants with high-quality fundus images were identified as suitable for further analysis. Among these, 5,125 patients with OCT data and 3,911 patients with fundus data experienced a CVD event. After excluding patients whose images were acquired post-CVD events and those diagnosed with ocular diseases or diabetes, the final cohort comprised 2,142 patients for OCT images and 1,652 patients for fundus images who suffered CVD (\texttt{CVD+}). Notably, 977 \texttt{CVD+} patients had both retinal modalities available. For the patients without CVD or control group (\texttt{CVD-}), 38,886 participants had OCT data, and 38,553 participants had fundus data. Of these, 21,758 participants had both retinal modalities available (the detailed flow chart is described in the supplementary material Fig.\ref{fig:flow_chart_chapter_4}). For a comprehensive breakdown of the CVD events included in this study and their respective patient distributions (refer to Figure \ref{fig:cvd_heatmaps} in the supplementary material).
\begin{figure}[t!]
    \centering
    \includegraphics[width=\textwidth]{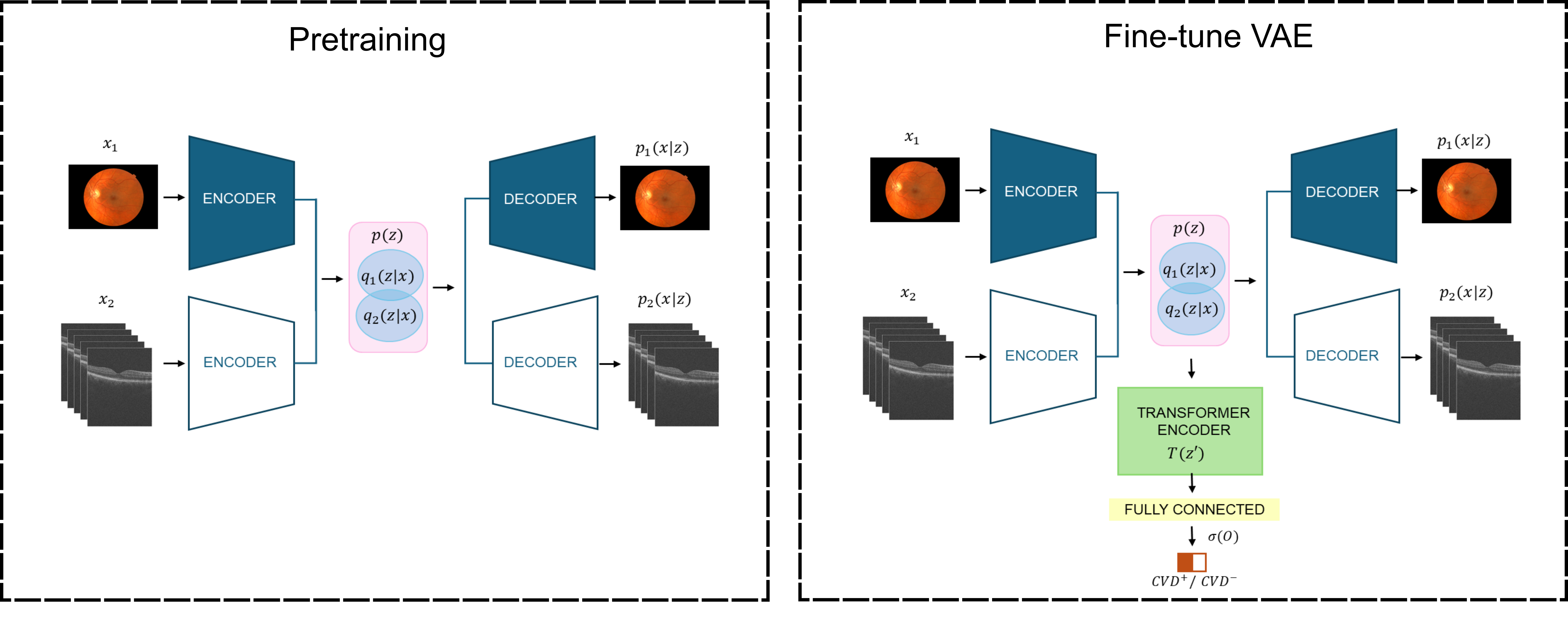}
    \caption{The illustration shows the architecture of our model. During the pre-training phase (left), fundus photographs and OCT images are processed through a 2D CNN and a 3D CNN encoder-decoder network in a multi-channel VAE configuration, respectively. In the fine-tuning phase (right), the images are processed through the pre-trained encoder-decoder networks. The resulting latent vectors are aggregated and input into a transformer architecture, followed by a fully connected layer. The model undergoes end-to-end training in both phases.}
    \label{fig:method}
\end{figure}
\subsection{Framework of Task-aware Multi-channel Variational Autoencoder (task-aware MCVAE)}
This study proposes a predictive model for classifying patients into \texttt{CVD+} and \texttt{CVD-}. The methodology comprises two stages: (1) training a Multi-channel Variational Autoencoder (MCVAE) \cite{Antelmi2019} using both retinal modalities as inputs from approximately 18,000 \texttt{CVD-} patients, with the objective of extracting features from retinal OCT and fundus images, and (2) implementing a transformer-based binary classifier. The framework inputs consist of volumetric OCT B-scans and fundus photographs, which are utilized to learn a compressed latent representation of the high-dimensional image data. This latent representation is subsequently employed by the transformer classifier to differentiate between \texttt{CVD-} subjects and those at risk of future CVD (\texttt{CVD+}) (see Fig. \ref{fig:method}).
\subsubsection{Pretraining Multi-Channel Variational Autoencoder.}
Consider a MCVAE with $M$ channels. For each channel $m$, let $x_m$ be the input data for channel $m$, $z$ the shared latent variable, and $\theta_m$ the parameters of the generative model for channel $m$. The joint probability distribution of the generative model is defined as
$[p(x_1, \ldots, x_M, z) = p(z) \prod_{m=1}^M p(x_m | z, \theta_m)]$ where $p(z)$ is typically a standard normal distribution $\mathcal{N}(0, I)$. The Evidence Lower Bound (ELBO) for the multi-channel VAE is:
\begin{equation}
\text{ELBO} = \mathbb{E}_{q(z|\mathbf{x},\phi)}\left[\sum_{m=1}^M \log p(x_m | z, \theta_m)\right] - \text{KL}(q(z|\mathbf{x},\phi) \| p(z))
\end{equation}
where $x={x_1, \ldots, x_M}$ and the first term represents the reconstruction term and encourages the model to accurately reconstruct the input data for each channel, and the second term, the KL divergence term regularizes the approximate posterior to be close to the prior distribution of $z$. Since the true posterior distribution $p(z | x_1, \ldots, x_M)$ is intractable, we approximate it with $q(z | x_1, \ldots, x_M, \phi)$, typically chosen as a multivariate Gaussian. The MCVAE is trained to minimize the ELBO, promoting accurate reconstruction of all channels while maintaining a regularized latent space. The shared latent representation $z$ captures common information across all channels, facilitating the joint modeling of heterogeneous data types \cite{Antelmi2019}.
\subsubsection{Transformer Encoder Classifier.}
To further utilize the latent representation $z$  obtained from the MCVAE, we introduce a classifier built using a transformer architecture \cite{vaswani2017attention}. The input to the classifier is the latent space representation $z$, which is fed into a series of transformer encoder layers. We propose using a transformer due to the Multi-Head Self-Attention mechanism has demonstrated the ability to focus on different parts of the input representation, capturing dependencies and interactions between different latent dimensions \cite{Benarab2022CNNTransEncAC}. The final output of the transformer encoder layers is passed through a fully connected layer followed by a softmax activation function to produce classification probabilities. The transformer encoder classifier is trained to minimize a binary cross-entropy loss function, enabling it to learn from the latent representations and make accurate predictions.
\subsubsection{Losses.}
The loss function used to train the proposed model is presented in equation \ref{loss_function}, and is composed of: (1) the mean square error (MSE) $\mathcal{L}_{MSE}$ loss (see eqn. \ref{mse_loss}) for the reconstruction error between the original data ($\mathbf{x}_{i} $) and reconstructed data ($\hat{\mathbf{x}_{i}}$); (2) the binary cross-entropy loss $\mathcal{L}_{BCE}$ (see eqn. \ref{bce_loss}) for the classification task, where, $y^g_i$ and $y^p_i$ denote the ground truth class label and predicted class probability for the i\textsuperscript{th} sample; and (3) the Kullback–Leibler divergence loss $\mathcal{L}_{KL}$ (see eqn. \ref{mse_loss}).
\begin{equation}\label{loss_function}
    \mathcal{L}_{task-aware-MCVAE} = \mathcal{L}_{recon} + \mathcal{L}_{class}
\end{equation}

\begin{equation}\label{mse_loss}
    \mathcal{L}_{recon} = \frac{1}{N} \sum_{i=1}^{N} \left( \mathbf{x}_{i} - \hat{\mathbf{x}}_{i} \right)^2 + \frac{1}{2} \sum_{i=1}^{N} \left[ 1 + \log \left( \sigma_{i}^{2} \right) - \sigma_{i}^{2} - \mu_{i}^{2} \right]
\end{equation}

\begin{equation}\label{bce_loss}
    \mathcal{L}_{class} = \mathcal{L}_{BCE} = -\frac{1}{N}\sum_{i=1}^{N} \left[ {y}^g_{i} * \log \left( {{y}^p_{i}} \right) + \left(  1 - {y}^g_{i} \right) * \log \left( 1 - {y}^p_{i} \right)\right]
\end{equation}
\subsubsection{Model Interpretation.} 
To elucidate the model's behavior and provide local explanations for its classification decisions, particularly in relation to the \texttt{CVD+} group, we employed a two-step approach leveraging both feature importance analysis and visualization techniques. Initially, we utilized the SHAP (SHapley Additive exPlanations) library to compute SHAP values \cite{Lundberg2017AUA}, enabling the identification of latent variables derived from both retinal imaging modalities that contribute most significantly to the CVD classifications. This approach allows for a quantitative assessment of feature importance within the learned latent space. Subsequently, we applied the methodology of optical flow maps, as developed in a previous study \cite{Garca2024PredictingRO}. This process focused on the latent vectors exhibiting the highest contribution to the classification outcomes. We systematically altered these critical latent vectors, visualized the resulting reconstructions of the modified latent representations, and subsequently calculated the optical flow between the original and perturbed reconstructions.

\section{Experiments Details}
The experimental procedures were conducted utilizing an NVIDIA Tesla M60 GPU. The model was implemented using PyTorch (v1.10.2), and optimal hyperparameters were determined through a grid search strategy employing five-fold cross-validation. The dataset was partitioned into training, validation, and test sets in a 5:2:3 ratio. For the fundus network, both encoder and decoder architectures comprised six 2D convolutional layers. The OCT image network utilized four 3D convolutional layers for both encoder and decoder components. In both instances, the encoder employed Rectified Linear Unit (ReLU) activations, while the decoder implemented Leaky Rectified Linear Unit (LeakyReLU) activations (refer to Figure \ref{fig:method}). 

Our end-to-end model architecture consists of two primary stages: the Multi-Channel Variational Autoencoder (MCVAE) training stage and the task-aware MCVAE transformer classifier training stage. The model accepts two types of retinal images as input: fundus photographs and B-scan OCT images. Each modality is processed by its respective encoder. The MCVAE is optimized to minimize the reconstruction loss (the discrepancy between original and reconstructed images) and the KL divergence between the latent variable distributions and the prior $p(z)$. In the fine-tuning phase, the latent representations z serve as input to the multimodal transformer classifier. The transformer encoder processes these latent representations, utilizing its capacity to capture complex dependencies and interactions within the latent space. It enhances the latent features through multi-head self-attention mechanisms and feed-forward neural networks, thereby learning richer representations. The transformer encoder's output is subsequently processed through a fully connected layer. During this stage, the model is trained to minimize the classification loss, employing a binary cross-entropy loss function that quantifies the difference between predicted and actual CVD classifications. All subjects' images used for pre-training the MCVAE were excluded from subsequent experiments, where the task-aware MCVAE was trained and evaluated to identify patients at risk of future CVD based on their retinal images. We employed an age-sex-matched cohort for the \texttt{CVD+} and \texttt{CVD-} groups. This approach has been demonstrated to offer a significant advantage in mitigating the influence of confounding variables on the predictive model \cite{Zhou2023AFM,Brown2022DetectingSL}. During the classification phase, we conducted a comprehensive investigation into the impact of latent representations derived from OCT and fundus imaging on the predictive task. This was accomplished by generating three datasets from the same cohort of around 3,000 patients, each comprising different combinations of data sources: (i) latent representations from fundus photographs; (ii) latent representations from OCT imaging; and (iii) latent representations from both fundus and OCT modalities.
\section{Results}
\subsubsection{Classification Performance.}
\begin{figure}
    \centering
    \includegraphics[width=\textwidth]{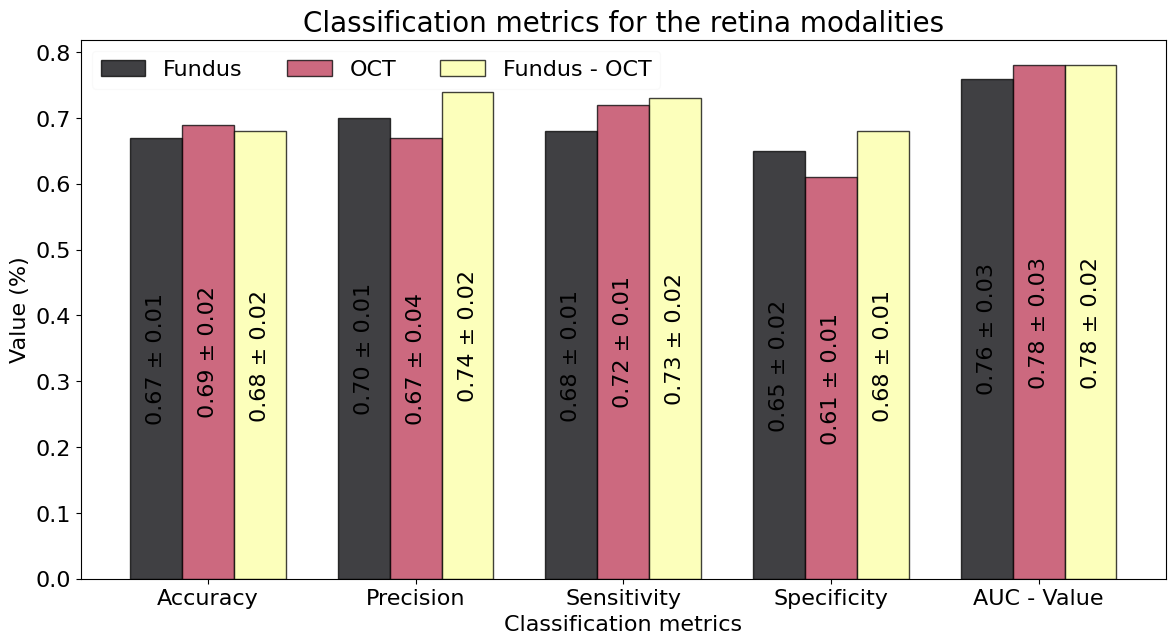}
    \caption{The bar chart displays the classification metrics for Fundus, OCT, and combined Fundus-OCT modalities. The metrics shown include Accuracy, Precision, Sensitivity, Specificity, and AUC (Area Under the Curve) values, with their respective standard deviations. The colors represent different modalities: Fundus (black), OCT (red), and Fundus-OCT (yellow). Each bar is labeled with its corresponding value.}
    \label{fig:metrics}
\end{figure}
In this study, we aim to evaluate the efficacy of two retinal imaging modalities—fundus photography and OCT— for predicting patients at risk of having a CVD event. To accomplish this objective, we have constructed datasets comprising three distinct cohorts: one utilizing solely fundus images, another employing only OCT images, and a third integrating both retinal modalities (Fundus-OCT). This methodological approach enables a comprehensive evaluation of each modality's contribution to the prediction task and facilitates an assessment of whether the integration of both modalities enhances predictive performance. To rigorously evaluate the model's efficacy, we have employed the three classifiers on an identical, previously unseen test set, comprising 1,000 \texttt{CVD-} subjects and 100 \texttt{CVD+} subjects. Our primary aim is to investigate whether the amalgamation of latent features derived from OCT images and fundus photographs provides superior discriminative power compared to the utilization of either modality in isolation. 

\begin{table}[h]
\centering
\begin{tabular}{|c|c|c|c|}
\hline
\textbf{Metric} & \textbf{Fundus CI} & \textbf{OCT CI} & \textbf{Fundus-OCT CI} \\ \hline
Accuracy & (0.662, 0.677) & (0.6827, 0.697) & (0.675, 0.687) \\ \hline
Precision & (0.693, 0.707) & (0.662, 0.678) & (0.733, 0.746) \\ \hline
Sensitivity & (0.674, 0.688) & (0.713, 0.727) & (0.723, 0.738) \\ \hline
Specificity & (0.641, 0.658) & (0.606, 0.619) & (0.672, 0.687) \\ \hline
AUC & (0.753, 0.767) & (0.772, 0.787) & (0.773, 0.788) \\ \hline
\end{tabular}

\caption{95\% Confidence Intervals (CI) for classification metrics of the Fundus, OCT, and Fundus-OCT classifiers. Each interval represents the range within which the true metric value is expected to lie with 95\% confidence, providing insight into the precision and reliability of the classifiers' performance.}
\label{tab:ci_metrics}
\end{table}

The results indicate that the combined Fundus-OCT classifier generally demonstrates superior or equivalent performance to the individual Fundus and OCT classifiers across the majority of metrics, with particular emphasis on precision and sensitivity. The observed minor variations in accuracy and Area Under the Curve (AUC) among the modalities fall within the margins of error, suggesting that while the integration of modalities tends to enhance performance, these improvements may not always achieve statistical significance. Nevertheless, the combined Fundus-OCT modality consistently exhibits higher or equal performance across various metrics, lending support to the hypothesis that the integration of multiple retinal imaging modalities can provide more comprehensive and discriminative features for CVD prediction. The overlapping confidence intervals suggest that the differences in performance metrics among the modalities are not highly significant (refer to Table \ref{tab:ci_metrics}). However, the consistent performance improvement observed with the combined modality (Fundus-OCT) indicates practical benefits in the simultaneous utilization of both imaging modalities.

\subsubsection{Ablation Study.}
\begin{table}
\centering
\begin{tabular}{|c|c|c|c|c|c|c|}
\hline
\textbf{Metrics} & \textbf{Fundus} & \textbf{Fundus CI} & \textbf{OCT} & \textbf{OCT CI} & \textbf{Fundus - OCT} & \textbf{Fundus - OCT CI} \\
\hline
Accuracy & 0.58 & 0.56 - 0.60 & 0.62 & 0.60 - 0.64 & 0.60 & 0.58 - 0.62 \\
\hline
Precision & 0.60 & 0.58 - 0.62 & 0.57 & 0.54 - 0.60 & 0.62 & 0.59 - 0.65 \\
\hline
Sensitivity & 0.58 & 0.56 - 0.60 & 0.56 & 0.52 - 0.60 & 0.62 & 0.59 - 0.65 \\
\hline
Specificity & 0.56 & 0.53 - 0.59 & 0.58 & 0.55 - 0.61 & 0.55 & 0.52 - 0.58 \\
\hline
AUC - Value & 0.60 & 0.57 - 0.63 & 0.63 & 0.60 - 0.66 & 0.64 & 0.63 - 0.65 \\
\hline
\end{tabular}
\caption{Classification metrics for the retina modalities (Ablation Study). This table compares the performance of classifiers trained without the end-to-end approach, highlighting the reduced efficacy across all metrics when the MCVAE is not trained alongside the classifier.}
\label{table:classification_metrics}
\end{table}
To evaluate the efficacy of our end-to-end training approach, we conducted an ablation study in which only the classifier component was trained, omitting the training of the MCVAE. In this experimental scenario, our loss function was restricted to the classification loss exclusively (refer to eqn \ref{bce_loss}).

Consistent with our previous findings, the utilization of both retinal modalities yielded superior outcomes across nearly all classification metrics. The results demonstrate that, when compared to our comprehensive end-to-end model, the classification metrics for all three classifiers (Fundus, OCT, and Fundus-OCT) were significantly diminished in the ablation study. This observed discrepancy suggests that the end-to-end training approach enhances the model's capacity to learn and focus on the salient features that distinguish \texttt{CVD+} and \texttt{CVD-} patients, thereby improving overall classification performance.
The classification metrics for the retinal modalities in the ablation study are presented in Table \ref{table:classification_metrics}. It is evident that while the Fundus-OCT classifier maintains its superior performance relative to the individual Fundus and OCT classifiers, the overall performance is attenuated compared to the end-to-end trained model. Specifically, we observed reductions in all classification metrics. These findings underscore the critical importance of comprehensive end-to-end training for optimizing classifier performance.

This ablation study provides valuable insights into the collaborative effects of joint feature learning and classification in our proposed model architecture. The observed performance degradation in the absence of MCVAE training suggests that the end-to-end approach facilitates the extraction of more relevant and discriminative features from the retinal imaging data. This, in turn, contributes to enhanced classification accuracy and robustness.
\subsubsection{Interpretation.}
\begin{figure}
    \centering
    \includegraphics[width=\textwidth]{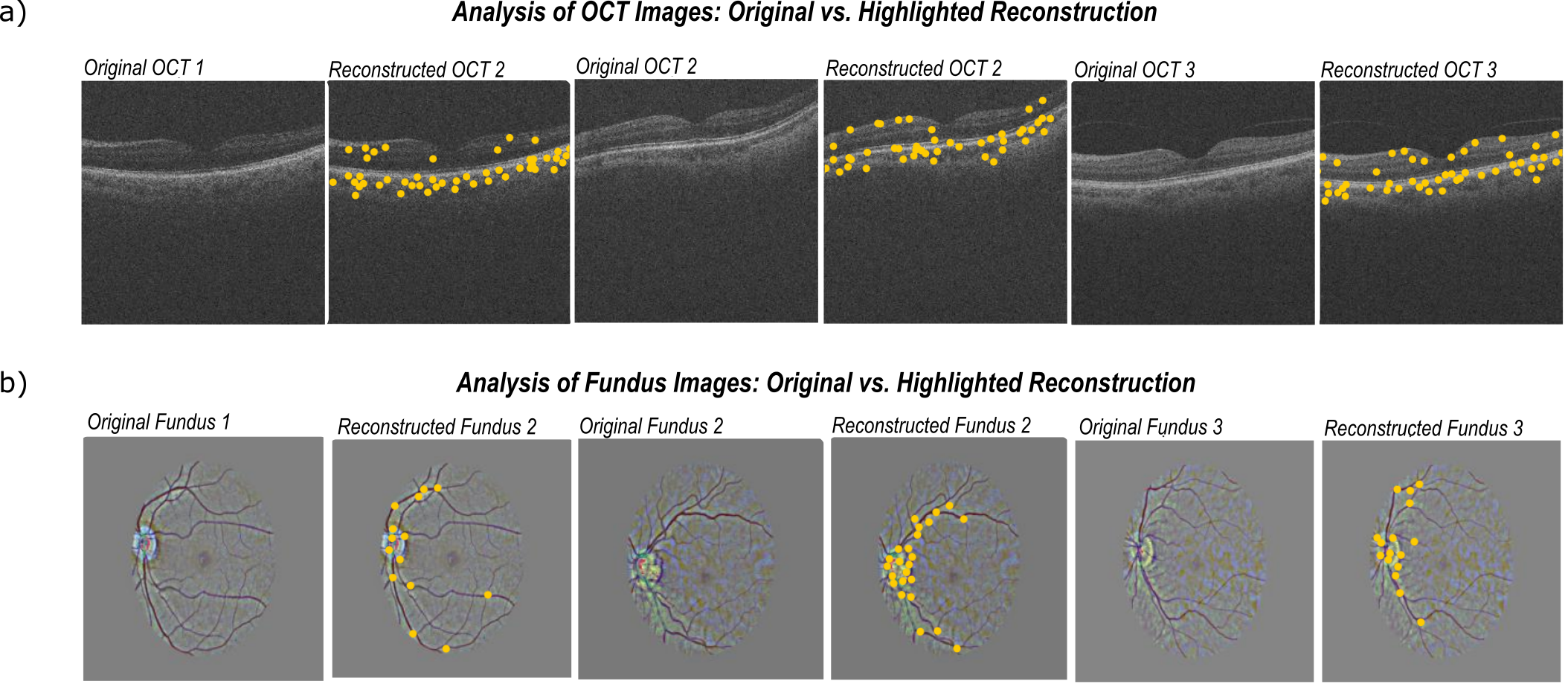}
    \caption{ a) The top row displays a sequence of Optical Coherence Tomography (OCT) images, highlighting the different layers of the retina with yellow markers indicating regions of interest. b) The bottom row shows corresponding fundus images with vascular structures and regions of interest also marked in yellow. These images are utilized for analyzing and predicting cardiovascular disease risks based on retinal biomarkers.}
    \label{fig:optical_flow}
\end{figure}

Following the identification of latent vectors exhibiting the highest contribution to the prediction of patients at elevated risk of future cardiac events using the SHAP values (see Figure \ref{fig:latent_vectors} in the supplementary material), we systematically altered the z values of both OCT and fundus images that demonstrated the highest values. For each case, we reconstructed the retinal modality utilizing the modified z values and subsequently estimated the optical flow maps between the original reconstructed image and the perturbed reconstructed image. In Figure \ref{fig:optical_flow}, we present examples from three patients. For simplicity, in the case of OCT, we only show the 64th B-scan example, although the entire volume was processed. Figure \ref{fig:optical_flow}(a) displays the qualitative OCT results, while Figure \ref{fig:optical_flow}(b) shows the fundus results, where the yellow points highlight the most pronounced differences. 
The optical flow maps generated by our model predominantly emphasized the choroidal layer in the OCT volume, as well as the retinal pigment epithelium (RPE). These results suggest that the choroidal layer may play a pivotal role in distinguishing between \texttt{CVD+} and \texttt{CVD+} patients.
In the case of fundus images, blood vessels were prominently highlighted in the majority of patients in the unseen test set, along with the optic disc. This observation aligns with established clinical indicators of retinal vascular health and its association with cardiovascular risk. These findings provide quantitative evidence of the specific retinal features that our model prioritizes in its classification process, offering insights into the potential biomarkers for cardiovascular risk assessment using multi-modal retinal imaging.
\section{Discussion}
Our findings demonstrate the efficacy of latent representations derived from retinal OCT images and fundus photographs, learned through a MCVAE, in predicting CVD when employed within a multimodal transformer classifier. This outcome corroborates previous studies that have identified relevant vascular biomarkers and cardiac health indicators within retinal imaging \cite{Garca2024PredictingRO,Zhou2023AFM}.  Our study utilized a pre-trained MCVAE that integrates B-scan OCT images and fundus photographs as distinct modalities, enabling our model to identify specific attributes within both types of retinal images that significantly contribute to CVD prediction. To our knowledge, only one other study has employed both fundus and OCT modalities for predicting certain CVDs; however, that study was limited to a single B-scan of the OCT, did not train the modalities jointly, and provided limited detail on the advantages and contributions of each modality \cite{Zhou2023AFM}.
A key contribution of our study is the interpretation of specific features in OCT images and fundus photographs that are relevant to the classification task, providing insights into the most discriminative regions of the retinal image. Our results suggest that choroidal morphology is a significant predictor of CVD risk, aligning with previous studies reporting associations between choroidal characteristics and the risk of stroke and acute myocardial infarction \cite{Yeung2020CHOROIDALTI}. Similarly, our model highlighted the optic disc and prominent veins in fundus images, which have been extensively studied and recognized as indicators of cardiovascular health. Abnormalities in retinal vessel caliber, branching patterns, and overall vascular geometry have been consistently correlated with CVD risk in previous studies. Furthermore, alterations in the morphology of the optic disc, including variations in its size, shape, and coloration, have been associated with various cardiovascular risk factors \cite{Guo2020AssociationBC}. These features are instrumental in elucidating the systemic impacts of cardiovascular health on the retinal microvasculature.
The concordance between our model's emphasis on these specific retinal features and established clinical indicators of cardiovascular risk provides validation for our approach. It suggests that our deep learning model has successfully learned to identify and prioritize clinically relevant features in both OCT and fundus images for cardiovascular risk prediction. This alignment between machine learning-derived features and known clinical markers not only enhances the interpretability of our model but also reinforces its potential clinical utility.
\paragraph{Future work:} Future research directions will aim to incorporate modalities from other organs to provide a more comprehensive context and additional information for treating CVD. The integration of multi-organ imaging data could potentially enhance the model's predictive accuracy and provide a more holistic view of cardiovascular health. Additionally, the inclusion of an external validation dataset will be crucial for assessing the generalizability of our model across diverse populations and clinical settings. This step is essential for establishing the robustness and clinical applicability of our approach in real-world scenarios.

\section{Conclusion}
In conclusion, our study provides significant insights into the use of retinal imaging for predicting cardiovascular diseases and highlights the importance of multi-modal approaches in improving predictive performance. The integration of OCT and fundus images has shown promising results, paving the way for future advancements in non-invasive cardiovascular health monitoring.
\section*{Acknowledgements}
This project was funded by the following institutions: The Royal Academy of Engineering INSILEX (grant no. CiET1819$\backslash{19}$) (A.F.F.), UKRI Frontier Research Guarantee INSILICO (grant no. EP/Y030494/1) (N.R. and A.F.F.) and Consejo Nacional de Humanidades Ciencias y Tecnologías (CONAHCYT) (scholarship no. 766588) (C.M.G). The NIHR Manchester Biomedical Research Centre also funds the work of A.F.F.



\newpage
\section*{Supplementary Material}

\begin{figure}[H]
    \centering
    \includegraphics[width=\textwidth]{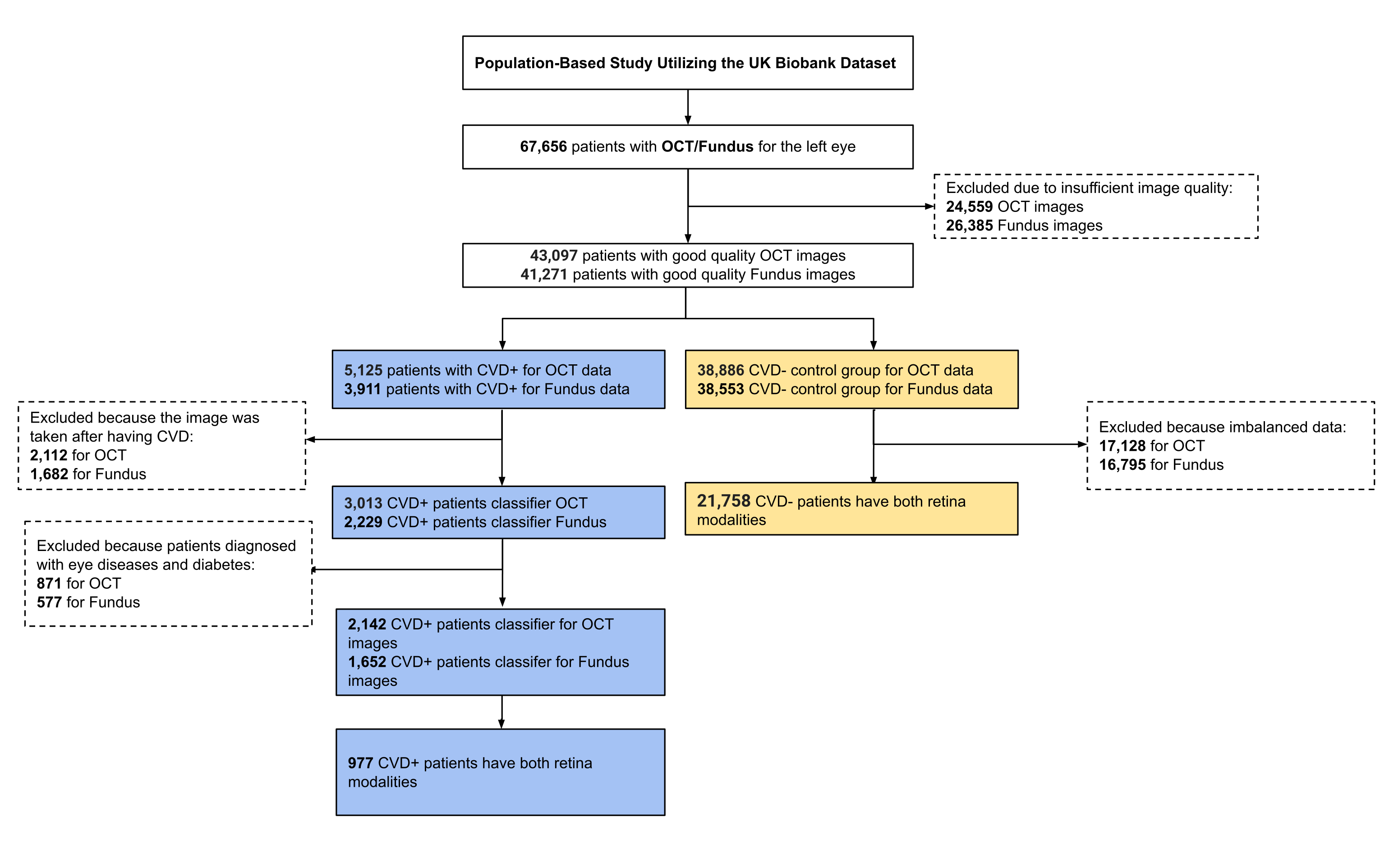}
    \caption{STROBE flow chart outlines the patient selection process for a study using the UK Biobank dataset with (Optical Coherence Tomography) OCT and Fundus images of the left eye.}
    \label{fig:flow_chart_chapter_4}
\end{figure}

Figure \ref{fig:flow_chart_chapter_4} provides a comprehensive outline of the inclusion and exclusion criteria for a population-based study utilizing the UK Biobank dataset. The focus of the study is on the left eye's OCT and Fundus images. This flow chart details a stringent selection process ensuring that only high-quality images are included in the final analysis. The study meticulously balances the cohorts by excluding cases with image quality issues, post-diagnosis images, other eye diseases, and imbalanced data, resulting in a robust dataset for assessing cardiovascular disease using retinal imaging.

\paragraph{\textbf{Heatmaps Depicting Patients with Different CVD.}}

\begin{figure}[H]
    \centering
    \begin{subfigure}[b]{0.8\textwidth}
        \centering
        \includegraphics[width=\textwidth]{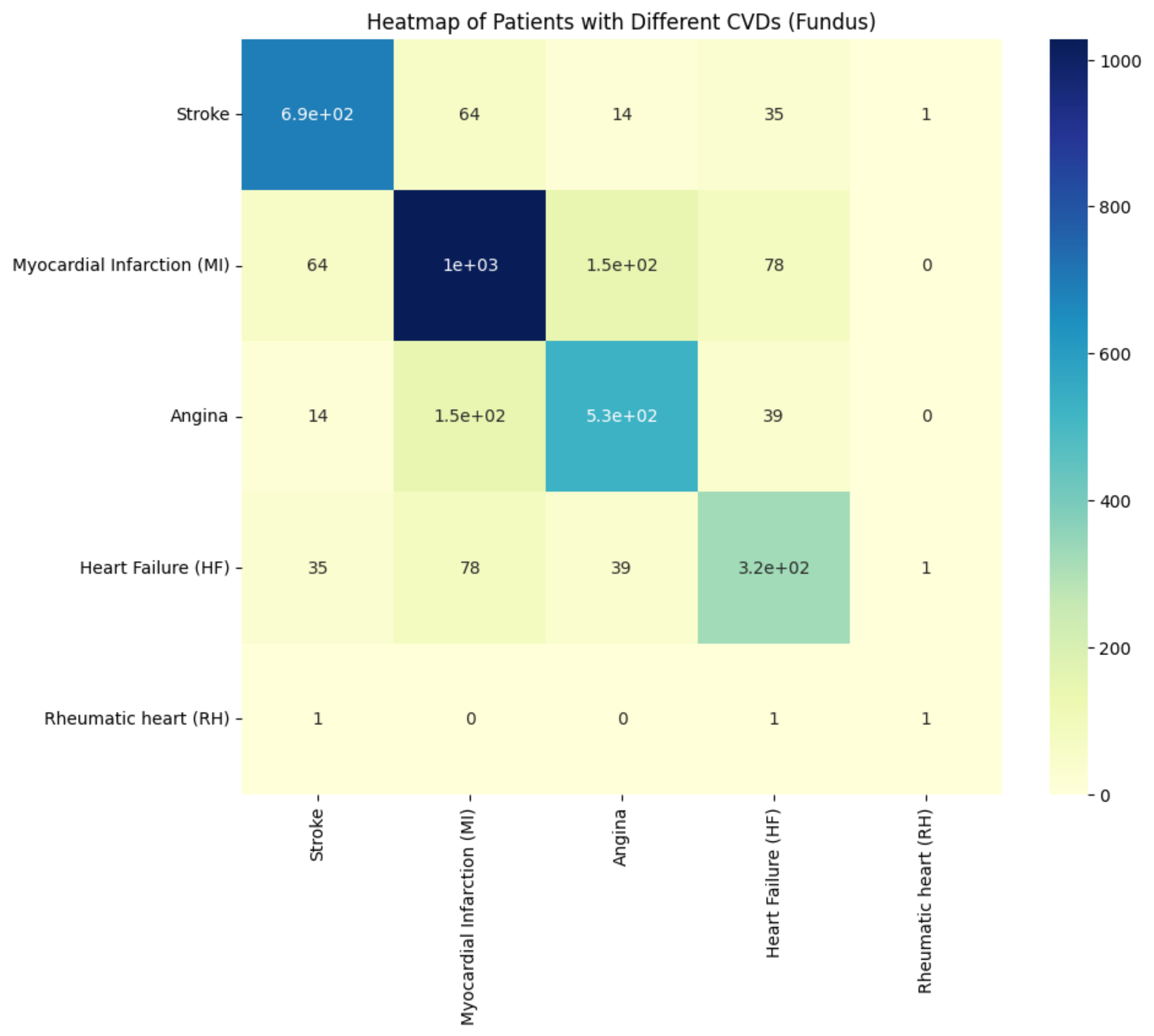}
        \caption{Fundus}
        \label{fig:cvd_heatmaps_fundus}
    \end{subfigure}
    \hfill
    \begin{subfigure}[b]{0.8\textwidth}
        \centering
        \includegraphics[width=\textwidth]{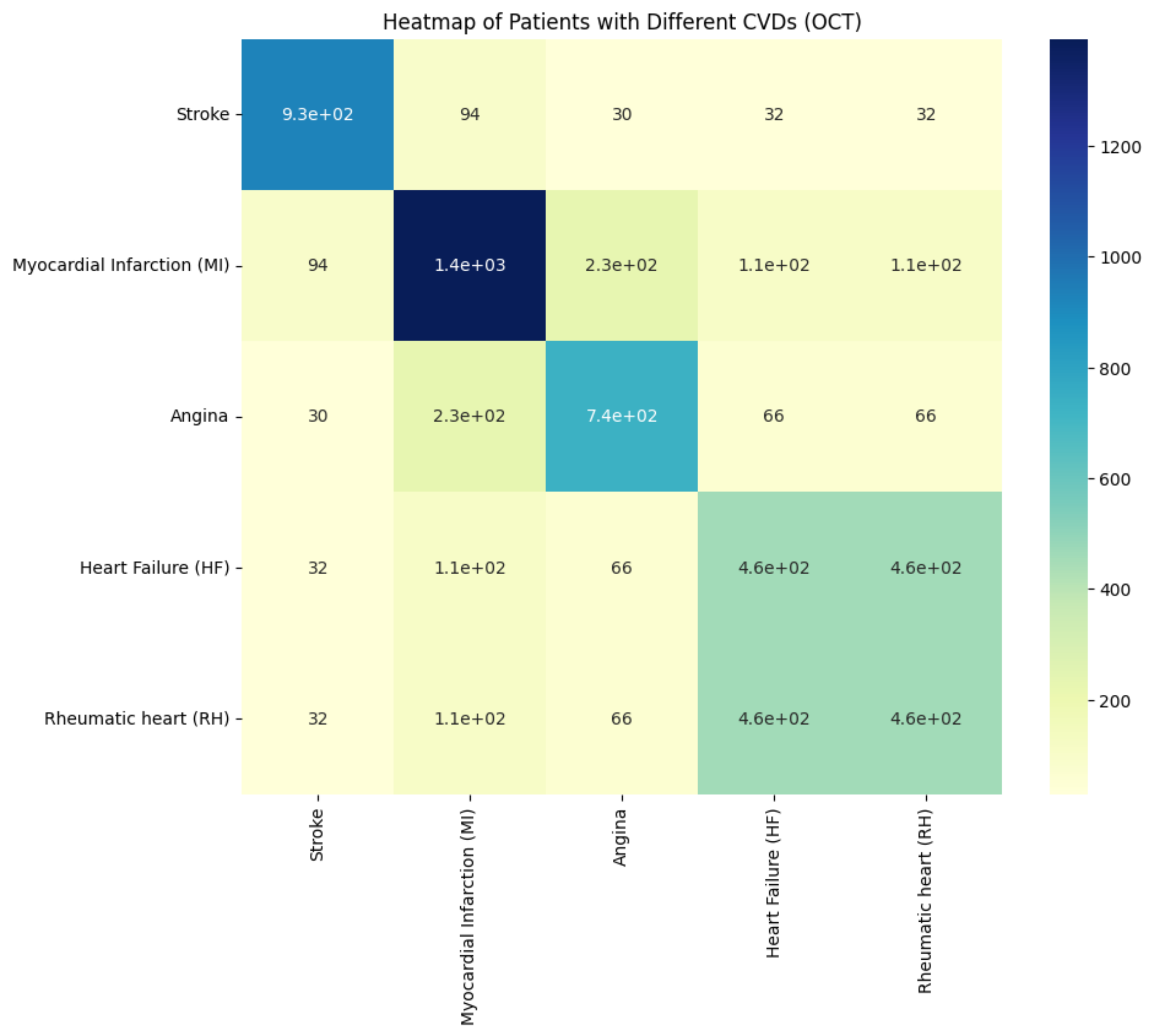}
        \caption{OCT}
        \label{fig:cvd_heatmaps_oct}
    \end{subfigure}
    \caption{Heatmaps of Patients with Different Cardiovascular Diseases (CVDs)}
    \label{fig:cvd_heatmaps}
\end{figure}

The provided heatmaps (refer to Figure \ref{fig:cvd_heatmaps})illustrate the distribution of patients with various CVDs across two different diagnostic imaging methods: Fundus and OCT. Each heatmap visually represents the number of patients diagnosed with specific CVDs.
\paragraph{\textbf{Fundus} (Figure \ref{fig:cvd_heatmaps_fundus})}: The highest concentration of stroke patients is observed, with a count of 690. MI has the highest number of patients, totaling 1000. It also shows some overlap with Stroke (64), Angina (150), and Heart Failure (78). There are 530 patients with Angina, with notable overlaps with MI (150) and Heart Failure (39). A significant number of patients (320) are reported with eart Failure (HF), with minor overlaps with other CVDs. Very few patients fall into this Rheumatic heart, with minimal overlap with other diseases.
\paragraph{\textbf{OCT} (Figure \ref{fig:cvd_heatmaps_oct})}: The highest concentration is 930 for Stroke, with minor overlaps with MI (94) and other CVDs. MI also shows the highest number of patients (1400), with overlaps in Angina (230) and HF (110). There are 740 patients with Angina, with significant overlaps with MI (230) and Heart Failure (66). A considerable number of patients (460) are recorded with HF, with overlaps in Angina (66) and Rheumatic Heart (32). Similar to the Fundus heatmap, RH shows a small number of patients, with minor overlaps with other diseases.

\paragraph{\textbf{Confusion matrix for the three classifier.}}

Confusion matrices for the classification of retinal images using Fundus, OCT, and Fundus-OCT modalities (see Table \ref{tab:confusion_matrix}). Each matrix displays the true positives (TP), false negatives (FN), false positives (FP), and true negatives (TN) for each modality, along with the totals for prediction positives, prediction negatives, and overall totals

\begin{table}
    \centering
    \begin{tabular}{|>{\columncolor{gray!20}}c|c|c|c|}
        \hline
        & Prediction Positive & Prediction Negative & Total \\
        \hline
        \rowcolor{gray!20} Fundus & & & \\
        Real Positive & TP (70) & FN (30) & 100 \\
        Real Negative & FP (30) & TN (650) & 1000 \\
        \hline
        Total & 100 & 1000 & 1100 \\
        \hline
        \rowcolor{gray!20} OCT & & & \\
        Real Positive & TP (67) & FN (33) & 100 \\
        Real Negative & FP (33) & TN (610) & 1000 \\
        \hline
        Total & 100 & 1000 & 1100 \\
        \hline
        \rowcolor{gray!20} Fundus-OCT & & & \\
        Real Positive & TP (73) & FN (27) & 100 \\
        Real Negative & FP (26) & TN (680) & 1000 \\
        \hline
        Total & 100 & 1000 & 1100 \\
        \hline
    \end{tabular}
    \caption{Confusion matrices for Fundus, OCT, and Fundus-OCT}
    \label{tab:confusion_matrix}
\end{table}

\paragraph{\textbf{Latent Vectors with the Most Significant Influence on the Task.}}
\begin{figure}[H]
    \centering
    \includegraphics[width=\textwidth]{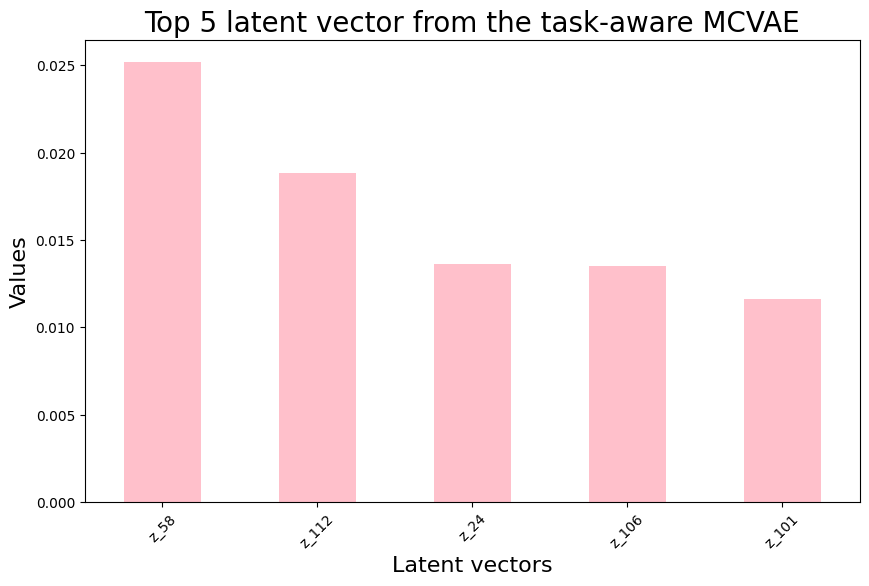}
    \caption{The top 5 latent vectors \( z_{58} \), \( z_{112} \), \( z_{24} \), \( z_{106} \), and \( z_{101} \) are identified based on their SHAP values, which measure the contribution of each feature to the model's prediction of cardiovascular disease risk.}
    \label{fig:latent_vectors}
\end{figure}

Figure \ref{fig:latent_vectors} illustrates the top 5 latent vectors that have the greatest influence on the prediction of CVD risk, as identified using the SHAP methodology. These vectors represent the features extracted by the task-aware MCVAE model that contribute most significantly to the predictive performance. Specifically, the vectors $z_{58}$, $z_{112}$, $z_{24}$, $z_{106}$, and $z_{101}$ are shown to have the highest SHAP values, indicating their importance in the model's decision-making process.

\end{document}